\documentclass[12pt]{article}
\usepackage{times}		
\usepackage{amsfonts}		%for fraktur font
\usepackage{amssymb}		%for \blacksquare symbol (proof terminator)
\usepackage{theorem}		%for setting body font of theorems
\usepackage{cite}		%for \citen command
\usepackage{array}		%for increasing row spacing in arrays
\theorembodyfont{\upshape}
\newtheorem{theorem}{Theorem}[section]
\newtheorem{proposition}[theorem]{Proposition}
\newtheorem{lemma}[theorem]{Lemma}
\newtheorem{formula}[theorem]{Formula}
\newtheorem{corollary}{Corollary}[theorem]

\newcommand{\defeq}{\stackrel{\rm{def}}{=}}

\begin{document}

\title{\bf On Crossing Event Formulas in Critical Two-Dimensional Percolation}
\author{Robert S. Maier\footnote{Departments of Mathematics and Physics,
University of Arizona, Tucson, Arizona 85721;  
email: rsm@math.arizona.edu; phone: +1 520 621 2617; fax: +1 520 621 8322.}}
\date{}
\maketitle

\begin{abstract}
Several formulas for crossing functions arising in the continuum limit of
critical two-dimensional percolation models are studied.  These include
Watts's formula for the horizontal-vertical crossing probability and
Cardy's new formula for the expected number of crossing clusters.  It~is
shown that for lattices where conformal invariance holds, they simplify
when the spatial domain is taken to be the interior of an equilateral
triangle.  The two crossing functions can be expressed in~terms of an
equianharmonic elliptic function with a triangular rotational symmetry.
This suggests that rigorous proofs of Watts's formula and Cardy's new
formula will be easiest to construct if the underlying lattice is
triangular.  The simplification in a triangular domain of Schramm's `bulk
Cardy's formula' is also studied.
\end{abstract}


\eject

\section{Introduction}

The critical behavior of percolation is not fully understood, either
rigorously or formally.  As~the percolation threshold is approached,
connected clusters occur with high probability on ever larger length
scales.  It~has been conjectured that in any dimension, there is a
universal scaling limit of isotropic short-range percolation models,
defined over the continuum and independent of the details of the model,
such as the lattice and percolation type (site or~bond)~\cite{Aizenman96}.
This continuum theory would capture the connectivity of typical
configurations of the underlying discrete model, at or near criticality.

Conformal field theory makes predictions for the crossing probabilities of
the continuum limit of critical two-dimensional percolation, and
especially, predicts that they are conformally invariant~\cite{Cardy2001}.
For example, if the continuum theory is confined to a spatial
domain~$\Omega$ plus boundary~$\partial\Omega$, and $\gamma_1,\gamma_2$ are
disjoint pieces of~$\partial\Omega$, the probability of the event that
$\gamma_1,\gamma_2$ are connected by a percolation cluster is predicted to
be invariant under transformations that are conformal on~$\Omega$ (though
not necessarily on~$\partial\Omega$).  The probabilities of more
complicated crossing events, involving more than two pieces
of~$\partial\Omega$, are also predicted to be invariant.

In~the first applications of conformal field theory to percolation,
$\Omega$~was taken to be a rectangle, with aspect ratio $r\defeq{\rm
width/height}$.  In~this geometry, Cardy~\cite{Cardy92} derived a formula
for the crossing function $\Pi_h(r)$, the probability that the two vertical
sides are horizontally connected.  His formula takes~on a simpler form if
the rectangle is conformally mapped onto the upper half plane
$\mathbb{H}\subset\mathbb{C}$, and its boundary to
$\partial\mathbb{H}=\mathbb{R}\cup\{\infty\}$.  The vertical sides are
mapped to disjoint line segments on the real axis, one of which can be
taken without loss of generality to be semi-infinite.  They are usually
taken to be $[0,z]$ and $[1,\infty]$, where $z\in(0,1)$, with $z=0,1/2,1$
corresponding to~$r=\infty,1,0$.  We~write
$\mathfrak{P}_h(z)\defeq\Pi_h\left(r(z)\right)$.  If~the underlying
discrete model is bond percolation on a square lattice, duality suggests
$\Pi_h(1/r)=1-\Pi_h(r)$, i.e., $\mathfrak{P}_h({1-z})=1-\mathfrak{P}_h(z)$,
so that $\Pi_h(1)=\mathfrak{P}_h(1/2)$, the probability that two opposite
sides of a large square are connected by a critical percolation cluster,
should equal~$1/2$.

On the numerical side, crossing events and their conformal invariance were
extensively investigated by Langlands and
collaborators~\cite{Langlands92,Langlands94}, and it was verified that
Cardy's formula is valid for discrete percolation models on a rectangular
square lattice of size $L\times L'$, with $r=L/L'$, in the limit
$L,L'\to\infty$.  They also investigated $\Pi_{hv}(r)$, the probability
that all four sides of the rectangle are connected.  Watts~\cite{Watts96}
derived a formula for the equivalent function
$\mathfrak{P}_{hv}(z)\defeq\Pi_{hv}\left(r(z)\right)$ from conformal field
theory by making additional assumptions, and his formula agrees well with
the data of Langlands et~al.  It~should be noted that
$\mathfrak{P}_{hv}\le\mathfrak{P}_h$, and that by symmetry,
$\Pi_{hv}(1/r)=\Pi_{hv}(r)$, i.e.,
$\mathfrak{P}_{hv}(1-z)=\mathfrak{P}_{hv}(z)$.  Also, it~is clear that
$\Pi_{hv}(r)/\Pi_h(r)\to1$ as~$r\to\infty$, i.e.,
$\mathfrak{P}_{hv}(z)/\mathfrak{P}_h(z)\to1$ as~$z\to0$.

Cardy~\cite{Cardy2000,Cardy2001} later derived a formula for the expected
number of percolation clusters that cross between $\gamma_1,\gamma_2$
(the~left and right-hand sides of a rectangle, or the two corresponding
line segments in~$\partial\mathbb{H}$).  This may be viewed as a function
$N_h(r)$, or equivalently a function $\mathfrak{N}_h(z)\defeq
N_h\left(r(z)\right)$; necessarily, $\mathfrak{N}_h\geq \mathfrak{P}_h$.

So in all, three crossing formulas for critical percolation have been
derived from conformal field theory: formulas for $\mathfrak{P}_h(z)$,
$\mathfrak{P}_{hv}(z)$, and $\mathfrak{N}_h(z)$, which yield formulas for
$\Pi_h(r)$, $\Pi_{hv}(r)$, and $N_h(r)$.  Recently,
Smirnov~\cite{Smirnov2001a} provided the first rigorous proof of any of
these, namely Cardy's formula for~$\mathfrak{P}_h(z)$.  He~showed that
critical site percolation on the triangular lattice has a conformally
invariant scaling limit, and that discrete percolation cluster boundaries
converge to a stochastic Loewner evolution process.  These facts followed
from his proof that in a general domain~$\Omega$, a~conformally transformed
Cardy's formula is valid.  This includes a version due~to L.~Carleson
(unpublished, but see ref.~\citen{Cardy2001}, sec.~7.2, and
ref.~\citen{Werner2000}, prediction~7).  Suppose $\mathbb{H}$~is mapped
conformally onto an equilateral triangle~$\triangle ABC$ in the complex
plane, and that the map extends to the boundary, with $[-\infty,0]$,
$[0,1]$, $[1,\infty]$ being mapped to the edges $AB$, $BC$,~$CA$.  So $w\in
BC$ and $z\in(0,1)$ will correspond.  Carleson noticed that
$\widetilde{\mathfrak{P}}_h(w)\defeq\mathfrak{P}_h\left(z(w)\right)$ simply
equals $\overline{Bw}/\overline{BC}$.  That~is, Cardy's formula is formally
equivalent to the statement that that the line segment~$Bw$ is connected by
a critical percolation cluster to the opposite side of the triangle,~$CA$,
with probability equal to the fraction of the side~$BC$ occupied by~$Bw$.
Equivalently, the real-valued function $\widetilde{\mathfrak{P}}_h(w)$~is
the restriction to one side of the triangle of a particularly simple
analytic function of~$w$: a~{\em linear\/} function.

The triangular symmetry of Carleson's restatement made possible Smirnov's
proof, which is specific to a triangular lattice.  Smirnov notes,
%\begin{quote}
``It~seems that $2\pi/3$ rotational symmetry enters in our paper not
because of the specific lattice we consider, but rather [because~it]
manifests some symmetry laws characteristic to (continuum) percolation.''
%\end{quote}
Whether his proof extends to other lattices is unclear.

In this paper we study whether the predicted formulas for the functions
$\mathfrak{P}_{hv}(z)$ and~$\mathfrak{N}_h(z)$, like Cardy's formula
for~$\mathfrak{P}_{h}(z)$, simplify when the spatial domain~$\Omega$ is
taken to be an equilateral triangle, rather than a rectangle or the upper
half plane~$\mathbb{H}$.  We~show that they~do.  In~particular, we show
that the four-way crossing function
$\widetilde{\mathfrak{P}}_{hv}(w)\defeq\mathfrak{P}_{hv}\left(z(w)\right)$
predicted for the equilateral triangle has its second derivative
$\widetilde{\mathfrak{P}}''_{hv}(w)$ equal to a familiar elliptic function:
an~equianharmonic Weierstrass $\wp$-function, where `equianharmonic'
signifies that the period lattice of the $\wp$-function is triangular, with
a $\pi/3$ rotational symmetry.  (See ref.~\citen{Abramowitz65}, sec.~18.13.)
This contrasts with the linear function $\widetilde{\mathfrak{P}}_h(w)$,
the second derivative of which is zero.  Our new representation for
$\widetilde{\mathfrak{P}}_{hv}(w)$, i.e., for $\mathfrak{P}_{hv}(z)$
or~$\Pi_{hv}(r)$, immediately yields a simple closed-form expression for
$\mathfrak{P}_{hv}(1/2)=\Pi_{hv}(1)$, the probability that all four sides
of a large square are connected by a critical percolation cluster; namely,
$1/4+(\sqrt3/4\pi)(3\log3-4\log2)\approx 0.322$.  It~would be difficult
though not impossible to derive this expression directly from Watts's
formula.

We show that Cardy's recent formula for~$\mathfrak{N}_h(z)$, like Watts's
formula for $\mathfrak{P}_{hv}(z)$, simplifies in an equilateral triangle.
$\widetilde{\mathfrak{N}}''_{h}(w)$, the second derivative of
$\widetilde{\mathfrak{N}}_{h}(w)\defeq\mathfrak{N}\left(z(w)\right)$, can
also be expressed in~terms of the equianharmonic $\wp$-function.  In~fact,
we derive a curious identity relating the three crossing functions
$\widetilde{\mathfrak{P}}_{h}$, $\widetilde{\mathfrak{P}}_{hv}$,
and~$\widetilde{\mathfrak{N}}_h$, or equivalently $\mathfrak{P}_{h}$,
$\mathfrak{P}_{hv}$, and~$\mathfrak{N}_h$; namely, that
$2\mathfrak{N}_h(z)-\mathfrak{P}_h(z)-\mathfrak{P}_{hv}(z)$ must equal
$(\sqrt3/2\pi)\log\left(1/(1-z)\right)$.  Setting $z=1/2$ yields that
$\mathfrak{N}_h(1/2)$, i.e., $N_h(1)$, the expected number of critical
percolation clusters crossing between two opposite sides of a large square,
should equal $3/8+(\sqrt3/8\pi)(3\log3-2\log2)\approx 0.507$.  In~a final
study of a crossing formula, we treat a fourth Cardy-type formula proved
rigorously by Schramm~\cite{Schramm2001} for triangular-lattice site
percolation, which is valid `in~the bulk' and does not follow from boundary
conformal field theory.  We~show it has a simple restatement in a suitable
triangular domain.

Our successful simplification of the crossing event formulas indicates that
an equilateral triangular domain is a good `fit' to the continuum limit of
critical percolation.  It~also suggests that rigorous proofs of Watts's
formula and Cardy's new formula will be easiest to construct if the
underlying lattice is triangular.  Our restatements contrast with those of
Ziff~\cite{Ziff95,Ziff95a} and Kleban and
Zagier~\cite{Kleban2000,Kleban2002}, which also involve higher
transcendental functions.  They focused on the continuum limit of
percolation in a rectangle, and especially on the derivatives $\Pi_h'(r)$
and~$\Pi_{hv}'(r)$ of the rectangular crossing functions.  In~our notation,
Ziff showed that $\Pi_h'(r)$ is proportional to $[\theta_1'(0,q=e^{-\pi
r})]^{4/3}$, where $\theta_1(\cdot,q)$~is the first Jacobi theta function.
Also, Kleban and Zagier showed that $\Pi_h',\Pi_{hv}'$, considered jointly,
have interesting modular transformation properties, and that these
properties characterize $\Pi_{hv}'(r)$.  The relation between our results
and theirs is not yet clear.

Section~\ref{sec:formulas} presents each crossing event formula in a
standard form.  Section~\ref{sec:math} covers conformal mapping concepts,
including the equianharmonic $\wp$-function.  The simplified versions of
the crossing formulas that apply in triangular domains are derived in
Section~\ref{sec:restatements}.  An appendix reviews some basic
mathematical facts.

\section{Crossing Event Formulas}
\label{sec:formulas}

The four crossing formulas for continuum percolation on the closed upper
half plane~$\overline{\mathbb{H}}$ involve hypergeometric functions, both
Gauss's ${}_2F_1$ and Clausen's ${}_3F_2$.  They can be stated in a
standardized, P-symbol form.  (For hypergeometric functions and P-symbols,
see the appendix.)  For the first three formulas, $\partial\mathbb{H}$~is
divided into $[-\infty,0]$, $[0,z]$, $[z,1]$, and~$[1,\infty]$.
A~connection between $[0,z]$ and $[1,\infty]$ corresponds to a horizontal
crossing on the original rectangle, and one between $[-\infty,0]$
and~$[z,1]$ to a vertical crossing.  The probability of a horizontal
connection is~$\mathfrak{P}_h(z)$, with $\mathfrak{N}_h(z)$ the expected
number of such connections.  All four segments are connected with
probability $\mathfrak{P}_{hv}(z)$.  Let $\mathfrak{P}_{h\bar
v}\defeq\mathfrak{P}_h-\mathfrak{P}_{hv}$, the probability of there being a
horizontal connection that is not also a vertical one.

\begin{formula}[{\rm Cardy~\cite{Cardy92}}]
\label{form:1}
The function $\mathfrak{P}_h(z)$ equals
\begin{equation}
\frac{3\,\Gamma(2/3)}{\Gamma(1/3)^2}\,
z^{1/3} {}_2F_1
\biggl(
\begin{array}{c}
{1/3,\,2/3} \\
{4/3}\\
\end{array}
\!
\biggm|
z\,
\biggr)
\propto
P\left\{
\begin{array}{ccc|c}
0 & 1 & \infty & z \\
\hline
0 & 0 & 0  & \\
\fbox{\rm 1/3} & 1/3 & 1/3  & \\
\end{array}
\right\}.
\end{equation}
\end{formula}

\begin{formula}[{\rm Watts~\cite{Watts96}}]
\label{form:2}
The function $\mathfrak{P}_{h\bar v}(z)$ equals
\begin{equation}
\frac{\sqrt3}{2\pi}
\,z\, {}_3F_2
\biggl(
\begin{array}{c}
{1,\,1,\,4/3} \\
{2,\,5/3}\\
\end{array}
\!
\biggm|
z\,
\biggr)
\propto
P\left\{
\begin{array}{ccc|c}
0 & 1 & \infty & z \\
\hline
0 & 0 & 0  & \\
1/3 & 1/3 & 1/3  & \\
\fbox{\rm 1} & 1 & 0  & \\
\end{array}
\right\}.
\end{equation}
\end{formula}

\begin{formula}[{\rm Cardy~\cite{Cardy2000,Cardy2001}}]
\label{form:3}
The function $\mathfrak{N}_{h}(z)$ equals
\begin{equation}
\label{eq:newnewCardy}
\frac12
-
\frac{\sqrt3}{4\pi}
\left[
\log(1-z)
+
(1-z)\,
{}_3F_2
\biggl(
\begin{array}{c}
{1,\,1,\,4/3} \\
{2,\,5/3}\\
\end{array}
\!
\biggm|
1-z\,
\biggr)
\right].
\end{equation}
\end{formula}

\noindent
{\it Proof.}
This version of the formula for~$\mathfrak{N}_h(z)$ is not well known.  The
version deduced by Cardy from boundary conformal field theory was
\begin{equation}
\frac12 -\frac{\sqrt3}{4\pi} \left[ \log(1-z) +2\sum_{m=1}^\infty
\frac{(1/3)_m}{(2/3)_m}\, \frac{(1-z)^m}{m} \right].
\end{equation}
Formula~\ref{form:3} follows from the series
representation~(\ref{eq:hyperseries}), if the $1/m$ factor in the summand
is written as $(1)_{m-1}(1)_{m-1}/(2)_{m-1}(m-1)!$.
\hbox{}\hfill$\blacksquare$

\begin{corollary}
\label{cor:1}
$2\mathfrak{N}_h(z)-\mathfrak{P}_h(z)-\mathfrak{P}_{hv}(z)=(\sqrt3/2\pi)\log\left(1/(1-z)\right)$.
\end{corollary}

\noindent
{\it Proof.}
This follows by combining Formulas \ref{form:1}--\ref{form:3}, with the aid
of the symmetry relations $\mathfrak{P}_h({1-z})=1-\mathfrak{P}_h(z)$ and
$\mathfrak{P}_{hv}({1-z})=\mathfrak{P}_{hv}(z)$.
\hbox{}\hfill$\blacksquare$

\medskip
Corollary~\ref{cor:1} ties $\mathfrak{N}_h$ to $\mathfrak{P}_h$ and
$\mathfrak{P}_{hv}$ in quite a strong way.  To~explain how, we~must sketch
the heuristic origins of Formula~\ref{form:2} in boundary conformal field
theory.  Watts was led to Formula~\ref{form:2} by considering ODEs
satisfied by correlation functions of boundary operators.  His candidate
for an ODE satisfied by~$\mathfrak{P}_{hv}$ was the fifth-order Fuchsian
equation
\begin{equation}
\label{eq:fifth}
\left\{\frac{d^3}{dz^3} [z(z-1)]^{4/3} \frac{d}{dz}
[z(z-1)]^{2/3} \frac{d}{dz}\right\} F = 0,
\end{equation}
which has P-symbol
\begin{equation}
\label{eq:fifthP}
P\left\{
\begin{array}{ccc|c}
0 & 1 & \infty & z \\
\hline
0 & 0 & 0  & \\
1/3 & 1/3 & 1/3  & \\
0 & 0 & 0  & \\
1 & 1 & 1  & \\
2 & 2 & 2  & \\
\end{array}
\right\},
\end{equation}
but is not of hypergeometric type.  Due~to a factorization of the
differential operator on the left-hand side of this
equation~\cite{Kleban2000,Kleban2002}, its solution space properly contains
the solution space of the third-order equation
\begin{equation}
\label{eq:third}
\left\{\frac{d}{dz} [z(z-1)]^{1/3} \frac{d}{dz}
[z(z-1)]^{2/3} \frac{d}{dz}\right\} F = 0,
\end{equation}
which has P-symbol
\begin{equation}
\label{eq:thirdP}
P\left\{
\begin{array}{ccc|c}
0 & 1 & \infty & z \\
\hline
0 & 0 & 0  & \\
1/3 & 1/3 & 1/3  & \\
1 & 1 & 0  & \\
\end{array}
\right\},
\end{equation}
and is of hypergeometric type.  Furthermore, the solution space
of~(\ref{eq:third}) properly contains the solution space of the
second-order equation with P-symbol
\begin{equation}
\label{eq:secondP}
P\left\{
\begin{array}{ccc|c}
0 & 1 & \infty & z \\
\hline
0 & 0 & 0  & \\
1/3 & 1/3 & 1/3  & \\
\end{array}
\right\},
\end{equation}
including the function $\mathfrak{P}_h$ of Formula~\ref{form:1}.  Watts
noticed that the solution space of~(\ref{eq:third}) includes a
one-dimensional subspace of functions equal to zero at $z=0$ and invariant
under $z\mapsto1-z$, which are criteria for the function
$\mathfrak{P}_{hv}$.  For this reason, he expected $\mathfrak{P}_{hv}$, and
$\mathfrak{P}_{h\bar v} = \mathfrak{P}_h - \mathfrak{P}_{hv}$ too, to be
solutions of~(\ref{eq:third}), as~well as of~(\ref{eq:fifth}).  This
insight led to Formula~\ref{form:2}, which incorporates the
P-symbol~(\ref{eq:thirdP}).

Kleban and Zagier~\cite{Kleban2002} noticed that the five-dimensional
solution space of~(\ref{eq:fifth}) is spanned by the solution space
of~(\ref{eq:third}), to which $\mathfrak{P}_{hv}$ and~$\mathfrak{P}_h$
belong, and the functions $\log z$ and $\log(1-z)$.  But by
Corollary~\ref{cor:1}, the function~$\mathfrak{N}_h$ is a linear
combination of $\mathfrak{P}_{hv}$, $\mathfrak{P}_h$, and~$\log(1-z)$,
implying the following mysterious fact.

\begin{corollary}
The function $\mathfrak{N}_h(z)$, like the functions $\mathfrak{P}_{hv}(z)$
and~$\mathfrak{P}_{h}(z)$, is a solution of Watts's fifth-order
differential equation, Eq.~(\ref{eq:fifth}).
\end{corollary}

For the fourth crossing formula, $\partial\mathbb{H}$~is divided into
$[-\infty,z]$ and~$[z,\infty]$, with $z\in\mathbb{R}$ unrestricted.
A~special boundary condition is imposed: on the underlying discrete
lattice, percolation along the line segment $[-\infty,z]$ is allowed
by~fiat.  Let a distinguished point in~$\mathbb{H}$ be chosen; without loss
of generality, let it be~${i}=\sqrt{-1}$.  Then the function
$\mathfrak{P}_{\it surr}(z)$ is defined to be the probability that $i$~is
surrounded by the percolation hull of~$[-\infty,z]$, i.e., the outermost
boundary of the percolation cluster in~$\mathbb H$ containing (``growing
from'')~$[-\infty,z]$.  One expects $\mathfrak{P}_{\it
surr}(-z)=1-\mathfrak{P}_{\it surr}(z)$, i.e., that $\mathfrak{P}_{\it
surr}(z)-1/2$ is odd in~$z$.

\begin{formula}[{\rm Schramm~\cite{Schramm2001}}]
\label{form:4}
The function $\mathfrak{P}_{\it surr}(z)$ equals $1/2$ plus
\begin{displaymath}
\frac{\Gamma(2/3)}{\sqrt\pi\,\Gamma(1/6)}\,
z\,{}_2 F_1
\biggl(
\begin{array}{c}
{1/2,\,2/3} \\
{3/2}\\
\end{array}
\!
\biggm|
-z^2\,
\biggr)
\propto
P\left\{
\begin{array}{ccc|c}
0 & 1 & \infty & -z^2 \\
\hline
0 & 0 & 0  & \\
\fbox{\rm 1/2} & 1/3 & 1/6  & \\
\end{array}
\right\}.
\end{displaymath}
In the P-symbol expression on the right, $z<0$ and $z>0$ correspond to
different branches, which are negatives of each other.
\end{formula}

\section{Conformal Maps and Fuchsian ODEs}
\label{sec:math}

The upper half plane $\mathbb H$~and any triangle $\triangle ABC$ without
boundary are homeomorphic as complex manifolds.  In~fact, by the
Poincar\'e--Koebe uniformization theorem, the complex plane~$\mathbb C$,
the Riemann sphere~$\mathbb{CP}^1\defeq\mathbb{C}\cup\{\infty\}$,
and~$\mathbb H$ are the only simply connected one-dimensional complex
manifolds, up~to conformal equivalence~\cite{Beardon84}.  To~transfer the
Fuchsian ODEs satisfied by the crossing functions from $\mathbb H$
to~$\triangle ABC$, an explicit expression for the conformal map $w=s(z)$,
i.e., $s:\mathbb{H}\to\triangle ABC$, is useful.  $s$~is a Schwarz triangle
function, defined by the Schwarz--Christoffel formula~\cite{Sansone69}.
It~extends to a map from the closure of~$\mathbb H$ in~$\mathbb{CP}^1$,
i.e., $\mathbb{H}\cup\mathbb{R}\cup\{\infty\}$, to the triangle with
boundary.  The conditions $s(0)=B$, $s(1)=C$, $s(\infty)=A$ uniquely
determine~$s$.

The inverse Schwarz function $S:\triangle ABC\to\mathbb H$ often has a
deeper significance than~$s$ does, as the equilateral triangle case
illustrates.  Equilateral triangles tile the plane, and the inverse
function extends to a function $S:\mathbb{C}\to\mathbb{CP}^1$ that maps
alternating triangles in a checkerboard fashion to the upper and lower
half planes.  In~modern treatments, this extended map is classed as one of
a handful of `universal' branched covers of~$\mathbb{CP}^1$ by $\mathbb C$,
$\mathbb{CP}^1$, or~$\mathbb H$ (see ref.~\citen{Serre92}, sec.~6.4).

Since the triangular lattice is doubly periodic, one expects that $S$~is an
elliptic function, and can equally well be viewed as a function on a
complex elliptic curve $\mathbb{C}/{\cal L}$, where ${\cal L}\defeq
2\omega\mathbb{Z}+2\omega'\mathbb{Z}$ is an appropriate lattice of periods
(the factors of~$2$ are traditional).  This is correct, as we briefly
sketch; for details, see Abramowitz and Stegun\cite{Abramowitz65}, sec.\
18.13, and Sansone and Gerretsen~\cite{Sansone69}, secs.\ 14.2--14.3.  The
canonical elliptic function on~$\mathbb{C}$ is the Weierstrass function
$\wp(w;g_2,g_3)$, defined as the solution of $(\wp')^2=f(\wp)\defeq
4\wp^3-g_2\wp - g_3$ with a unit-strength double pole at~$w=0$.  The
parameters $g_2,g_3\in\mathbb{C}$, both of which cannot be zero, are
related in a nontrivial way to the fundamental
half-periods~$\omega,\omega'$ of~$\wp$.  The {\em equianharmonic\/} case,
which has special symmetries, is the case when ${g_2=0}$.  Due~to the
homogeneity relation $\wp(w;g_2,g_3)=t^{2}\wp(tw;t^{-4}g_2,t^{-6}g_3)$, in
the equianharmonic case all nonzero~$g_3$ are equivalent, so henceforth
$g_3\in\mathbb{R}\setminus\{0\}$, in particular $g_3=1$, will be taken.
In~this case the fundamental half-periods $\omega,\omega'$ can be chosen to
be a complex conjugate pair with $\Re\omega>0$, $\Im\omega<0$.  If the
basic real half-period $\omega+\omega'>0$ is denoted~$\omega_2$, then
$\omega,\omega'$ will equal $(\frac12\mp\frac{\sqrt{-3}}2)\omega_2$.
So~the period lattice~$\cal L$ will be a triangular lattice, with a $\pi/3$
rotational symmetry about the origin.  Explicitly,
$\omega_2=\Gamma(1/3)^3/4\pi\approx1.530$.

The elliptic curve $\mathbb{C}/{\cal L}$ is homeomorphic to a torus and can
be viewed as the parallelogram with vertices $0$,~$2\omega$,
$2\omega_2$,~$2\omega'$, equipped with periodic boundary conditions.  The
$\wp$-function maps this parallelogram doubly onto~$\mathbb{CP}^1$.  Also,
$\wp'$~maps it triply onto~$\mathbb{CP}^1$.  It turns~out that in the
equianharmonic case, the torus, i.e., this period parallelogram, can be
subdivided into six equilateral triangles, each mapped by~$\wp'$ with unit
multiplicity onto the left or right half plane.  (See
ref.~\citen{Abramowitz65}, fig.~18.11, which is unfortunately not quite to
scale.)  Due~to this, the equilateral inverse Schwarz function~$S$ can be
chosen to be essentially~$\wp'$.  The map
\begin{equation}
z=S(w)\defeq 1/2 + \wp'(w)/2i
\end{equation}
will take the equilateral triangle $\triangle ABC\defeq\triangle
0\overline{W_0}{W_0}$ in $w$-space to the upper half plane in $z$-space,
where $W_0\defeq (1+\frac{\sqrt{-3}}3)\omega_2$.  It~will take the boundary
of $\triangle ABC$ to~$\mathbb{R}\cup\{\infty\}$, and the vertices
$w=A,B,C$ respectively to $z=\infty,0,1$.  As a map from $\mathbb{C}$
to~$\mathbb{CP}^1$, it will take alternating triangles to the upper and
lower half planes.  Each of these equilateral triangles, which tile the
plane, has one vertex in each of the congruence classes
$A+\mathcal{L},B+\mathcal{L},C+\mathcal{L}$, i.e., in each of the sets
$S^{-1}(\infty)=\mathcal{L},S^{-1}(0),S^{-1}(1)$.  These classes will be
denoted $[A],[B],[C]$.

The ODEs on $\mathbb{CP}^1\supset\mathbb{H}$ that are satisfied by the
functions of Formulas \ref{form:1}--\ref{form:3} can be pulled back to ODEs
on~$\mathbb{C}$ via the extended map $S:\mathbb{C}\to\mathbb{CP}^1$.  The
important thing to note when performing the pullback is that $S$~is a {\em
branched\/} cover of $\mathbb{CP}^1$ by~$\mathbb{C}$, the critical points
of which are the points in $[A]$, $[B]$, and~$[C]$.  At~any $w_0\in\mathbb
C$, necessarily $S(w)\sim S(w_0)+{\rm const}\times({w-w_0})^p$ to leading
order, where $p$~is the multiplicity with which $w_0$~is mapped
to~$S(w_0)$.  (If $S(w_0)=\infty$, the right-hand side must be replaced by
${\rm const}\times({w-w_0})^{-p}$.)  Each critical point has~$p=3$.

The pulled-back ODEs are tightly constrained by the following lemma, which
specifies how P-symbols are pulled back.  It~is proved by considering the
local (Frobenius) solutions at each point.
\begin{lemma}
\label{lemma:1}
Let $R:M\to\mathbb{CP}^1$ be a holomorphic map of one-dimensional complex
manifolds.  Consider $Lu=0$, an $n$th-order Fuchsian ODE
on~$\mathbb{CP}^1$.  It~can be pulled back via~$R$ to a Fuchsian ODE
$\tilde L\tilde u=0$ on~$M$, in the sense that if $u=u(z)$ satisfies $Lu=0$
then $\tilde u=\tilde u(w)\defeq u(R(w))$ will satisfy $\tilde L \tilde
u=0$.  The $n$~characteristic exponents of~$\tilde L$ at each $w_0\in M$
will equal those of~$L$ at $z_0=R(w_0)$, multiplied by the multiplicity
with which $w_0$~is mapped to~$z_0$.
\end{lemma}

As an illustration of the use of this lemma in pulling back ODEs, we give a
new proof of Whipple's quadratic transformation formula for~${}_3F_2$.
(The formula originally appeared in ref.~\citen{Whipple27}, with a
combinatorial proof; a simpler combinatorial proof is due to
Bailey~\cite{Bailey28}.  For a useful discussion placing the formula in
context, see Askey~\cite{Askey94}, but note the misprint in eq.~2.19.)
This new proof resembles Riemann's concise P-symbol proof of Kummer's
quadratic transformation formulas for~${}_2F_1$, which is summarized in
ref.~\citen{Andrews99}, sec.~3.9.  It~relies on the
expression~(\ref{eq:hyperPsymbol}) for the P-symbol associated to
${}_{q+1}F_q$, which is apparently not well known.
\begin{proposition}
\label{prop:Whipple}
Let $a,b,c\in\mathbb C$ with neither $a-b+1$ nor $a-c+1$ equal to
a~nonpositive integer.  Then
\begin{eqnarray*}
&&{}_3F_2
\biggl(
\begin{array}{c}
{a,\,b,\,c} \\
{a-b+1,\,a-c+1}\\
\end{array}
\!
\biggm|
w\,
\biggr)\\
&&\qquad
= (1-w)^{-a}\,
{}_3F_2
\biggl(
\begin{array}{c}
{a-b-c+1,\,a/2,\,(a+1)/2} \\
{a-b+1,\,a-c+1}\\
\end{array}
\!
\biggm|
\frac{-4w}{(1-w)^2}\,
\biggr)
\end{eqnarray*}
holds in a neighborhood of~$w=0$.
\end{proposition}
{\it Remark.} The two sides are defined for all~$w$ such that
$\left|w\right|<1$, resp.\ for all~$w$ such that
$\left|-4w/(1-w)^2\right|<1$.  So by analytic continuation, equality holds
at all points within the loop of the curve
$\left|4w\right|=\left|1-w\right|^2$ surrounding the origin.

\medskip
\noindent
{\it Proof.}  The functions of~$w$ on the two sides satisfy third-order
Fuchsian ODEs that follow from the $q=2$ case of~(\ref{eq:hyperODE}), the
ODE satisfied by ${}_{q+1}F_q$.  The left and right-hand functions are
determined by the additional condition that they be analytic at~$w=0$ and
equal unity there.  It~will therefore suffice to prove that the Fuchsian
ODEs corresponding to the two sides are the same up~to normalization, i.e.,
have the same solution spaces.  A~necessary condition for this is that
their P-symbols be the same, i.e., by the
representation~(\ref{eq:hyperPsymbol}), that
\begin{eqnarray}
\label{eq:long}
&&P\left\{
\begin{array}{ccc|c}
0 & 1 & \infty & w \\
\hline
0 & 0 & a  & \\
b-a & 1 & b  & \\
c-a & a-2b-2c+2 & c  & \\
\end{array}
\right\}\\
&&\qquad
=
(1-w)^{-a}\,
P\left\{
\begin{array}{ccc|c}
0 & 1 & \infty & R(w) \\
\hline
0 & 0 & a-b-c+1  & \\
b-a & 1 & a/2  & \\
c-a & 1/2 & (a+1)/2  & \\
\end{array}
\right\},
\nonumber
\end{eqnarray}
where $R:\mathbb{CP}^1\to\mathbb{CP}^1$ is defined by
$R(w)\defeq-4w/(1-w)^2$, or equivalently
\begin{eqnarray}
\label{eq:long2}
&&P\left\{
\begin{array}{ccc|c}
0 & 1 & \infty & w \\
\hline
0 & a & 0  & \\
b-a & a+1 & b-a  & \\
c-a & 2(a-b-c+1) & c-a  & \\
\end{array}
\right\}\\
&&\qquad=
P\left\{
\begin{array}{ccc|c}
0 & 1 & \infty & R(w) \\
\hline
0 & 0 & a-b-c+1  & \\
b-a & 1 & a/2  & \\
c-a & 1/2 & (a+1)/2  & \\
\end{array}
\right\}.
\nonumber
\end{eqnarray}
That is, $R$ must pull back the right-hand P-symbol in~(\ref{eq:long2}) to
the left-hand one.

The map $w\mapsto z\defeq R(w)$ takes $w=0,1,\infty$ to $z=0,\infty,0$
respectively, and also $w=-1$ to~$z=1$.  Its~critical points are $w=\pm1$,
each of which has double multiplicity.  The exponents in the columns
of~(\ref{eq:long2}) agree precisely with what Lemma~\ref{lemma:1} states:
the exponents of $w=0$ and $w=\infty$ are the same as those of~$z=0$, and
those of~$w=1$ are twice those of~$z=\infty$.  One might think the
left-hand P-symbol would have a fourth critical point, at~$w=-1$, with
exponents twice those of~$z=1$, i.e., $0,1,2$.  But as noted in the
appendix, those exponents are the signature of an ordinary point; so no
fourth column is present.

The preceding argument shows why the ${}_3F_2$ parameters of the
proposition take the values they do, but it does not quite prove the
proposition.  As~reviewed in the appendix, any Fuchsian ODE
on~$\mathbb{CP}^1$ that has three singular points and is of third order
(i.e., $q=2$) has $3q+2=8$ independent exponent parameters, which are
displayed in its P-symbol, and ${q\choose2}=1$ accessory parameter, which
is not.  For the two ODEs to be the same up~to normalization, they must
have the same P-symbol, and also the same accessory parameter.  The latter
is most readily verified by changing variables from $z=R(w)$ to~$w$ in the
right-hand ODE\null.  An~explicit computation, omitted here, shows that the
resulting pulled-back ODE is indeed the $q=2$ case of the
ODE~(\ref{eq:hyperODE}), with the parameters of the left-hand side.
\hbox{}\hfill$\blacksquare$

\medskip
What will be used in Section~\ref{sec:restatements} is the following
variant of Whipple's quadratic transformation.  It~seems not to have
appeared in the literature.
\begin{proposition}
\label{prop:Whipple2}
Let $a,b,c\in\mathbb C$ with $(a+b+c)/2$ not equal to
a~nonpositive integer.  Then
\begin{eqnarray*}
&&{}_3F_2
\biggl(
\begin{array}{c}
{a,\,b,\,c} \\
{2,\,(a+b+c)/2}\\
\end{array}
\!
\biggm|
w\,
\biggr)\\
&&\qquad
= (1-w)\,
{}_3F_2
\biggl(
\begin{array}{c}
{(a+1)/2,\,(b+1)/2,\,(c+1)/2} \\
{2,\,(a+b+c)/2}\\
\end{array}
\!
\biggm|
4w(1-w)\,
\biggr)
\end{eqnarray*}
holds in a neighborhood of~$w=0$, provided one of $a,b,c$ equals
unity.
\end{proposition}
{\it Remark.} The two sides are defined for all~$w$ such that
$\left|w\right|<1$, resp.\ for all~$w$ such that $\left|4w(1-w)\right|<1$.
So by analytic continuation, equality holds at all points that are both
within the circle $\left|w\right|=1$ and within the loop of the curve
$\left|4w(1-w)\right|=1$ surrounding the origin.

\medskip
\noindent
{\it Proof.}  This closely follows that of Proposition~\ref{prop:Whipple};
the details are left to the reader.  The only new feature is that equality
between the accessory parameters of the ODEs satisfied by the two sides
leads to an additional condition on the ${}_3F_2$ parameters, beside the
exponent conditions of Lemma~\ref{lemma:1}.  Changing variables from
$z=R(w)\defeq4w(1-w)$ to~$w$ in the right-hand ODE pulls it back to the
left-hand ODE, plus an extraneous term proportional to
$({a-1})({b-1})({c-1})$.  Provided one of $a,b,c$ equals unity, this
undesired term is absent.  \hbox{}\hfill$\blacksquare$

\section{Transformed and Restated Formulas}
\label{sec:restatements}

In this section we show how the crossing event formulas, Formulas
\ref{form:1}--\ref{form:4}, simplify in appropriately chosen triangular
domains.  The restated formulas appear in Propositions
\ref{prop:1}--\ref{prop:4}.  For all but Schramm's formula, the appropriate
triangle is equilateral.  The restatements of Cardy's formula and Schramm's
formula do not involve special functions.  The restatements of Watts's
formula and Cardy's new formula do involve elliptic functions, but elliptic
functions are significantly more familiar than Clausen's~${}_3F_2$.  The
restatement of Cardy's formula is of~course identical to Carleson's, but
the other three are new.

The Fuchsian ODEs on $\mathbb{CP}^1\supset\mathbb{H}$ of Formulas
\ref{form:1}--\ref{form:3} are pulled back to ODEs on
$\mathbb{C}\supset\triangle ABC$ via the inverse Schwarz function
$S:\mathbb{C}\to\mathbb{CP}^1$.  In~the normalization of the last section,
$z=S(w)$ equals $1/2+\wp'(w)/2i$, with $\wp$ the equianharmonic
$\wp$-function, satisfying $(\wp')^2=4\wp^3-1$.  The equilateral triangle
$\triangle ABC$ is $\triangle0\overline{W_0}W_0$, i.e., $\triangle0,\rho
e^{-i\pi/6},\rho e^{i\pi/6}$, with side length $\rho\defeq \left|W_0\right|
= 2\omega_2/\sqrt3$.  As~noted,
$\omega_2\defeq\Gamma(1/3)^3/4\pi\approx1.530$ is the basic real
half-period of~$\wp$.

Since $S$ maps $AB$, $BC$, $CA$ to $[-\infty,0]$, $[0,1]$, $[1,\infty]$
respectively, the line segment $BC=\overline{W_0}W_0$ is of primary
interest.  Its midpoint is $\omega_2=(\overline{W_0}+W_0)/2$, which is
mapped to~$1/2$.  As~a necessary preliminary, the behavior of the first few
antiderivatives of~$\wp'$ along~$BC$ will now be described.  (See the
tables in ref.~\citen{Abramowitz65}, sec.~18.13, where $W_0$~is
denoted~`$z_0$'.)  Relative to the midpoint, $\wp'$~is an odd function:
it~equals $-i$ at~$\overline{W_0}$ and $i$~at~$W_0$.  Its antiderivative
$\wp$ is even: it equals zero at $\overline{W_0}$ and~$W_0$, and $4^{-1/3}$
at the midpoint.  The negative antiderivative of~$\wp$ is the so-called
Weierstrass zeta function, plus an arbitrary constant.  The shifted
negative antiderivative $\zeta-\pi/2\sqrt3\omega_2$ is odd: it~equals
$i\pi/6\omega_2$ at~$\overline{W_0}$ and $-i\pi/6\omega_2$ at~$W_0$.  The
antiderivative of~$\zeta$ equals $\log\sigma$ plus an arbitrary constant,
where $\sigma$~is the Weierstrass sigma function, which equals
$e^{\pi/3\sqrt3}e^{-i\pi/6}$ at~$\overline{W_0}$ and
$e^{\pi/3\sqrt3}e^{i\pi/6}$ at~$W_0$; and $e^{\pi/4\sqrt3}2^{1/3}3^{-1/4}$
at the midpoint~$\omega_2$.  It~is easily checked that
\begin{equation}
\label{eq:double}
{}-{\rm Log}\,\sigma(w) + \frac{\pi}{2\sqrt3\,\omega_2}\,w - \frac{\pi}{6\sqrt3},
\end{equation}
which is a double antiderivative of~$\wp$, is an even function relative to
the midpoint: it~equals zero at $\overline{W_0}$ and~$W_0$, and
$(1/12)(\pi/\sqrt3 + 3\log 3 -4\log2)$ at the midpoint.  ${\rm Log}$
signifies the principal branch of the logarithm function.

\begin{proposition}[Cardy's formula, transformed; cf.~Carleson]
\label{prop:1}
If conformal invariance holds, Formula~\ref{form:1} corresponds on the
equilateral triangle~$\triangle ABC$ plus boundary to the following.
$\widetilde{\mathfrak{P}}_h(w)$, the probability that the boundary segments
$Bw$ and~$CA$ are connected by a percolation cluster, is the restriction
to~$BC$ of an analytic function that is {\em linear\/}.  Explicitly,
$\widetilde{\mathfrak{P}}_h(w)=(w-B)/(C-B)$.
\end{proposition}
{\it Proof.}  By Lemma~\ref{lemma:1}, the P-symbol of Formula~\ref{form:1}
is pulled back via~$z=S(w)$ to
\begin{equation}
P\left\{
\begin{array}{ccc|c}
[A] & [B] & [C] & w \\
\hline
0 & 0 & 0  & \\
1 & \fbox{\rm 1} & 1  & \\
\end{array}
\right\},
\end{equation}
where $[A],[B],[C]$ are the classes of points on the $w$-plane that are
mapped by~$S$ to $z=\infty,0,1$ respectively.  This is because these points
are the critical points of~$S$, and each has triple multiplicity.  But a
singular point with exponents~$0,1$ is effectively an ordinary point.  So
the pulled-back ODE on~$\mathbb{C}$ (in~particular, on~$\triangle ABC$) has
no~singular points and should be effectively
$(d^2/dw^2)\widetilde{\mathfrak P}_h(w)=0$, as can be verified by an
explicit computation.  \hbox{}\hfill$\blacksquare$

\begin{proposition}[Watts's formula, transformed]
\label{prop:2}
If conformal invariance holds, Formula~\ref{form:2} corresponds on the
equilateral triangle~$\triangle ABC$ plus boundary to the following.
$\widetilde{\mathfrak{P}}_{hv}(w)$, the probability that all four boundary
segments $AB$, $Bw$, $wC$, and~$CA$ are connected by a percolation cluster,
is the restriction to~$BC$ of an analytic function with the property that
the difference $\widetilde{\mathfrak{P}}_{h\bar v}(w) \defeq
\widetilde{\mathfrak{P}}_h(w) - \widetilde{\mathfrak{P}}_{hv}(w)$ is
proportional to $(w-B)^3$ as~$w\to B$, to~leading order.  Explicitly,
\begin{equation}
\label{eq:chief}
\widetilde{\mathfrak{P}}_{hv}(w) =
\,-\,\frac{3\sqrt3}{\pi}\,{\rm Log}\,\sigma(w)
+\frac{3}{2}\,\frac{w}{\omega_2} - \frac12\,,
\end{equation}
where $\sigma$~is the equianharmonic Weierstrass sigma function.
\end{proposition}
{\it Proof.}  By Lemma~\ref{lemma:1}, the P-symbol of Formula~\ref{form:1},
which partially specifies the ODE satisfied by ${\mathfrak{P}}_{hv}(z)$
and~${\mathfrak{P}}_{h}(z)$, is pulled back via~$z=S(w)$ to
\begin{equation}
\label{eq:newP}
P\left\{
\begin{array}{ccc|c}
[A] & [B] & [C] & w \\
\hline
0 & 0 & 0  & \\
1 & 1 & 1  & \\
0 & \fbox{\rm 3} & 3  & \\
\end{array}
\right\}.
\end{equation}
The condition $\Pi_{hv}(r)/\Pi_h(r)\to1$ as~$r\to\infty$, i.e.,
$\mathfrak{P}_{hv}(z)/\mathfrak{P}_h(z)\to1$ as~$z\to0$, implies
$\widetilde{\mathfrak{P}}_{hv}(w)/\widetilde{\mathfrak{P}}_h(w)\to1$
as~$w\to B$.  So~$\widetilde{\mathfrak{P}}_{hv}(w)$ is linear in~$w$
as~$w\to B$, to~leading order.  By the P-symbol~(\ref{eq:newP}), the first
nonzero correction must be cubic.

By changing variables from $z=S(w)$ to~$w$ in~(\ref{eq:third}), the
third-order ODE satisfied by ${\mathfrak{P}}_{hv}(z)$
and~${\mathfrak{P}}_{h}(z)$, one obtains the striking pulled-back ODE
\begin{equation}
\frac{d}{dw}\left\{
[\wp(w)]^{-1}\,\widetilde{\mathfrak{P}}''_{hv}(w)
\right\}=0
\end{equation}
on~$\mathbb{C}$.  So $\widetilde{\mathfrak{P}}_{hv}$~must be proportional
to a double antiderivative of~$\wp$.  The condition
$\Pi_{hv}(1/r)=\Pi_{hv}(r)$, i.e.,
$\mathfrak{P}_{hv}(1-z)=\mathfrak{P}_{hv}(z)$, implies that on the line
segment~$BC$, $\widetilde{\mathfrak{P}}_{hv}$~must be even around the
midpoint~$w=\omega_2$.  Moreover, the condition
$\widetilde{\mathfrak{P}}_{hv}(w)/\widetilde{\mathfrak{P}}_h(w)\to1$
as~$w\to B$ implies $\widetilde{\mathfrak{P}}_{hv}=0$ at the endpoints
$w=\overline{W_0},{W_0}$.  Any even double antiderivative of~$\wp$
equalling zero at $w=\overline{W_0},{W_0}$ must be a constant times the
function~(\ref{eq:double}).  For $\widetilde{\mathfrak{P}}_{hv}$, the
constant is set by
\begin{equation}
\widetilde{\mathfrak{P}}_{hv}'(B)= \widetilde{\mathfrak{P}}_{h}'(B)=1/(C-B)
=1/(W_0-\overline{W_0}) = \sqrt3/2i\omega_2,
\end{equation}
together with the fact that $[{\rm
Log}\,\sigma]'(B)=\zeta(\overline{W_0})$, the value of which is given
above.  By~examination, the constant should be~$3\sqrt3/\pi$; which
yields~(\ref{eq:chief}). 
\hbox{}\hfill$\blacksquare$

\medskip
\noindent
{\it Numerical Remark.}  A power series expansion of $\wp(w)$ about $w=W_0$
that is accurate to $O\left((w-W_0)^{15}\right)$ is given in
ref.~\citen{Abramowitz65}, eq.~18.13.41.  The corresponding expansion about
$w=\overline{W_0}$, i.e., about~$B$, is obtained by complex conjugation.
By twice anti-differentiating this, an expansion
of~$\widetilde{\mathfrak{P}}_{hv}(w)$ about $w=B$ accurate to
$O\left((w-B)^{17}\right)$ can be obtained.

\begin{corollary}
\label{cor:expression}
$\Pi_{hv}(1)$, the probability that all four sides of a large square are
connected by a critical percolation cluster, equals
\begin{equation}
\label{eq:expression}
1/4+(\sqrt3/4\pi)(3\log3-4\log2)\approx0.322.
\end{equation}
\end{corollary}
{\it Proof.}  $\Pi_{hv}(r=1)=\mathfrak{P}_{hv}(z=1/2)=
\widetilde{\mathfrak{P}}_{hv}(w=\omega_2)$ by conformal invariance.  This
quantity can be computed from~(\ref{eq:chief}), using the closed-form
expression for $\sigma(\omega_2)$ given at the beginning of this section.
\hbox{}\hfill$\blacksquare$

\medskip
\noindent
{\it Alternative Proof.} The expression~(\ref{eq:expression}) for
$\Pi_{hv}(1)$ can be derived directly from Watts's formula, though the
derivation is intricate; the following explains how. $\Pi_{hv}(1)$ equals
$\mathfrak{P}_{hv}(1/2)$, i.e., $\mathfrak{P}_{h}(1/2)-\mathfrak{P}_{h\bar
v}(1/2)$.  By Formula~\ref{form:2},
\begin{equation}
\label{eq:rewrite1}
\Pi_{hv}(1) = 1/2 -
\frac{\sqrt3}{4\pi}\,
{}_3F_2
\biggl(
\begin{array}{c}
{1,\,1,\,4/3} \\
{2,\,5/3}\\
\end{array}
\!
\biggm|
1/2\,
\biggr),
\end{equation}
since $\mathfrak{P}_h(1/2)=1/2$.  Summing the ${}_3F_2$~series requires
care, since in~general, it~is harder to evaluate ${}_3F_2$ than~${}_2F_1$.
For example, though Gauss's formula
\begin{equation}
{}_2F_1
\biggl(
\begin{array}{c}
{a,\,b} \\
{c}\\
\end{array}
\!
\biggm|
1\,
\biggr)
=\frac{\Gamma(c)\Gamma(c-a-b)}{\Gamma(c-a)\Gamma(c-b)},
\qquad \Re(c-a-b)>0,
\end{equation}
evaluates any ${}_2F_1$ at unit argument, no general formula for
${}_3F_2(a,b,c;d,e;1)$ in~terms of gamma functions
exists~\cite{Wimp83,Zeilberger92}.  However, certain special~${}_3F_2$'s
can be evaluated at unit argument in closed form.  At~other argument
values, the situation is unresolved.  Many `strange' evaluations of
${}_3F_2$ and~${}_2F_1$ at rational points other than unity are
known~\cite{Gessel82}, but most can apparently be reduced to evaluations at
unity via appropriate transformations of the independent variable.

To move the ${}_3F_2$ evaluation point in~(\ref{eq:rewrite1}) from $1/2$ to
unity, the new quadratic transformation formula of
Proposition~\ref{prop:Whipple2} can be used.  It yields
\begin{equation}
\label{eq:rewrite2}
\Pi_{hv}(1) = 1/2 -
\frac{\sqrt3}{8\pi}\,
{}_3F_2
\biggl(
\begin{array}{c}
{1,\,1,\,7/6} \\
{2,\,5/3}\\
\end{array}
\!
\biggm|
1\,
\biggr).
\end{equation}
(The point~$1/2$ is on the boundary of the region to which
Proposition~\ref{prop:Whipple2} applies; but by the $\Re
(\sum\beta_i-\sum\alpha_i) > 0$ convergence criterion mentioned in the
appendix, the equality of the proposition extends to the boundary.)
Fortunately, the~${}_3F_2(1)$ in~(\ref{eq:rewrite2}) can be evaluated in
closed form.  An ingenious application of L'H\^opital's rule to Gauss's
formula shows that
\begin{equation}
{}_3F_2
\biggl(
\begin{array}{c}
{1,\,1,\,a} \\
{2,\,c}\\
\end{array}
\!
\biggm|
1\,
\biggr)
=
\frac{c-1}{a-1}
\,\left[
\psi(c-1)-\psi(c-a)
\right],\quad
a\neq1,\ \Re(c-a)>0
\end{equation}
(see Luke~\cite{Luke75}, sec.~5.2.4).  Here $\psi\defeq\Gamma'/\Gamma$ is
the digamma function.  So
\begin{equation}
\Pi_{hv}(1) = 1/2 -\frac{\sqrt3}{2\pi}\,
\left[\psi(2/3)-\psi(1/2)\right].
\end{equation}
The values $\psi(2/3)$, $\psi(1/2)$ are $-\gamma+\pi/2\sqrt3-(3/2)\log 3$
and $-\gamma-2\log2$ respectively, where $\gamma$~is Euler's constant.
(See ref.~\citen{Prudnikov86}, vol.~2, app.~II.3.)  Substitution yields the
expression (\ref{eq:expression}) for~$\Pi_{hv}(1)$.
\hbox{}\hfill$\blacksquare$

\begin{proposition}[Cardy's new formula, transformed]
\label{prop:3}
If conformal invariance holds, Formula~\ref{form:3} corresponds on the
equilateral triangle~$\triangle ABC$ plus boundary to the following.
$\widetilde{\mathfrak{N}}_h(w)$, the expected number of percolation
clusters connecting the boundary segments $Bw$ and~$CA$, is the restriction
to~$BC$ of an analytic function.  Explicitly,
$\widetilde{\mathfrak{N}}_h(w)$ equals
\begin{displaymath}
{}-\frac{\sqrt3}{4\pi}
\left\{
6\,{\rm Log}\,\sigma(w) + {\rm Log}\,\left[\frac12-\frac{\wp'(w)}{2i}\right]
\right\}
+\frac{(3-\sqrt3\,i)}{4}\,\frac{w}{\omega_2} + \frac{\sqrt{3}\,i}{4}\,.
\end{displaymath}
\end{proposition}
{\it Proof.}  This follows from Corollary~\ref{cor:1} by replacing
${\mathfrak{N}}_h(z)$, ${\mathfrak{P}}_h(z)$, ${\mathfrak{P}}_{hv}(z)$,~$z$
by $\widetilde{\mathfrak{N}}_h(w)$, $\widetilde{\mathfrak{P}}_h(w)$,
$\widetilde{\mathfrak{P}}_{hv}(w)$,~$S(w)$, respectively, and substituting
the expressions for $\widetilde{\mathfrak{P}}_h(w)$,
$\widetilde{\mathfrak{P}}_{hv}(w)$ provided by Propositions
\ref{prop:1}~and~\ref{prop:2}.  \hbox{}\hfill$\blacksquare$

\begin{corollary}
\label{cor:expression2}
$N_{h}(1)$, the expected number of critical percolation clusters crossing
between opposite sides of a large square, equals
\begin{equation}
3/8+(\sqrt3/8\pi)(3\log3-2\log2)\approx 0.507.
\end{equation}
\end{corollary}
{\it Proof.}  $N_{h}(r=1)=\mathfrak{N}_{h}(z=1/2)=
\widetilde{\mathfrak{N}}_{h}(w=\omega_2)$ by conformal invariance.  This
quantity can be computed from the formula for
$\widetilde{\mathfrak{N}}_{h}(w)$, using the known value of
$\sigma(\omega_2)$ and the fact that $\wp'(\omega_2)=0$.  More simply, it
follows from Corollary~\ref{cor:1} by substituting the expression for
$\Pi_{hv}({r=1})= {\mathfrak{P}}_{hv}({z=1/2})$ provided by
Corollary~\ref{cor:expression}.  \hbox{}\hfill$\blacksquare$

\medskip
Finally we come to Schramm's formula, Formula~\ref{form:4}.  It differs
from Formulas \ref{form:1}--\ref{form:3} in~that its restatement employs a
triangular domain that is not equilateral.  Let $\triangle
A'B'C'\subset\mathbb{C}$ be an isosceles triangle with interior angles
$2\pi/3,\pi/6,\pi/6$ at $A',B',C'$, respectively.  (For concreteness, take
the vertices $A',B',C'$ equal to $1+i\sqrt3/3,0,2$, respectively.)  Special
boundary conditions are imposed: the edge~$B'C'$ is divided into $B'w$
and~$wC'$, and on the underlying discrete lattice, percolation along~$B'w$
is allowed by~fiat.  Also, for percolation purposes the edges $A'B'$
and~$A'C'$ are identified, so that in~effect, the boundary of~$\triangle
A'B'C'$ comprises only the edge~$B'C'$, and the vertex~$A'$ is in its
interior.

\begin{proposition}[Schramm's formula, transformed]
\label{prop:4}
If conformal invariance holds, Formula~\ref{form:4} corresponds on the
triangle $\triangle A'B'C'$ plus boundary, with edges $A'B'$, $A'C'$
identified, to the following.  Let $\widetilde{\mathfrak{P}}_{\it surr}(w)$
denote the probability that the vertex~$A'$ is surrounded by the
percolation hull of the boundary segment~$B'w$, i.e., the outermost
boundary of the percolation cluster growing from~$B'w$.  Then
$\widetilde{\mathfrak{P}}_{\it surr}(w)$ is the restriction to~$B'C'$ of an
analytic function that is {\em linear\/}.  Explicitly,
$\widetilde{\mathfrak{P}}_{\it surr}(w)=(w-B')/(C'-B')$.
\end{proposition}
{\it Proof.}  The first thing to observe is that up~to trivial changes of
the independent and dependent variables, the function $\mathfrak{P}_{\it
surr}(z)$ of Formula~\ref{form:4} is identical to the
function~$\mathfrak{P}_h(z)$ of Cardy's Formula~\ref{form:1}, or more
accurately to its analytic continuation.  This is the source of the linear
behavior on~$\triangle A'B'C'$.  To see the close relation between the two
functions, use Lemma~\ref{lemma:1} to pull back the P-symbol of
Formula~\ref{form:4} via the quadratic map $z\mapsto -z^2$
on~$\mathbb{CP}^1$.  The result of this procedure is that
$\mathfrak{P}_{\it surr}(z)$ equals $1/2$ plus a function in the solution
space specified by
\begin{equation}
\label{eq:quadratic}
P\left\{
\begin{array}{ccc|c}
0 & 1 & \infty & -z^2 \\
\hline
0 & 0 & 0  & \\
1/2 & 1/3 & 1/6  & \\
\end{array}
\right\}
=
P\left\{
\begin{array}{ccc|c}
-i & i & \infty & z \\
\hline
0 & 0 & 0  & \\
1/3 & 1/3 & 1/3  & \\
\end{array}
\right\},
\end{equation}
since the quadratic map has $0,\infty$ as its critical points (of
multiplicity~$2$), and takes $0,\pm i,\infty$ to $0,1,\infty$.  There is
no~fourth column associated to $z=0$ in the right-hand P-symbol, since in
the pulled-back ODE, $z=0$ has exponents $0,1$ and is effectively an
ordinary point.  The close connection between this P-symbol and the
P-symbol of Formula~\ref{form:1} is obvious.  A~careful computation,
omitted here, yields
\begin{equation}
\mathfrak{P}_{\it surr}(z) = 1/2
+{\rm const}\times\left[\mathfrak{P}_h(1/2+iz/2) - 1/2\right].
\end{equation}
But for the purpose of proving the proposition, (\ref{eq:quadratic})~will
suffice.

Via a conformal map~$R$ similar to the map~$S$ used in the proofs of
Propositions \ref{prop:1}--\ref{prop:3}, the right-hand P-symbol
in~(\ref{eq:quadratic}) can be pulled back to a trivial P-symbol.  However,
it~will turn~out that $R$~maps a triangle $\triangle A'B'C'$ of the above
form not onto~$\mathbb H$, but rather onto the {\em slit\/} half plane
$\mathbb H\setminus[i,+\infty i)$.  The edges $A'B'$, $A'C'$ will be mapped
to opposite sides of the slit, and will therefore need to be identified for
percolation purposes.  $A'$~will be mapped to~$i$, so the statements of the
proposition and Formula~\ref{form:4} will correspond.  The map $R$~and
$\triangle A'B'C'$ are chosen as~follows.

The function $S(w)=1/2+\wp'(w)/2i$ maps $\triangle ABC=\triangle
0,\overline{W_0},W_0$ onto~$\mathbb{H}$, and its vertices to
$\infty,0,1\in\partial\mathbb{H}=\mathbb{R}\cup\{\infty\}$.  So
$R(w)\defeq\wp'(w)$ maps $\triangle 0,\overline{W_0},W_0$ onto the
right half plane, and its vertices to $\infty i,-i,i\in
\mathbb{R}i\cup\{\infty i\}$.  By reflecting through the line passing
through $\overline{W_0}$ and~$W_0$ (and their midpoint, the real
half-period~$\omega_2$), it~follows that as~well, $R$~maps the triangle
$\triangle 2\omega_2,W_0,\overline{W_0}$ onto the left half plane.
Therefore $R$~maps the parallelogram without boundary
$0,\overline{W_0},2\omega_2,W_0$ comprising these two equilateral triangles
and the line segment $\overline{W_0}W_0$ (their common boundary) onto the
doubly slit plane $\mathbb{C}\setminus[i,+\infty i)\setminus(-\infty
i,-i]$.  In particular, it~maps the upper half of this parallelogram, the
isosceles triangle $\triangle W_0,0,2\omega_2$, onto the slit upper half
plane.  This triangle has interior angles $2\pi/3,\pi/6,\pi/6$.  Its
vertices are mapped to $i,+\infty i,+\infty i$, respectively.

One accordingly chooses $\triangle A'B'C'=\triangle W_0,0,2\omega_2 \propto
\triangle (1+i\sqrt3/3),0,2$, where the constant of proportionality
equals~$\omega_2$.  The pullback proceeds as in the proof of
Proposition~\ref{prop:1}.  The right-hand P-symbol of~(\ref{eq:quadratic})
is pulled back via $z=R(w)$ to
\begin{equation}
P\left\{
\begin{array}{ccc|c}
[A] & [B] & [C] & w \\
\hline
0 & 0 & 0  & \\
1 & 1 & 1  & \\
\end{array}
\right\}.  
\end{equation}
So the pulled-back ODE on $\triangle A'B'C'$ (or~more generally,
on~$\mathbb{C}$) has no~singular points and should be effectively
$(d^2/dw^2)\widetilde{\mathfrak P}_{\it surr}(w)=0$, as can be verified by
an explicit computation.  Therefore $\widetilde{\mathfrak P}_{\it surr}$
must be linear.  Since $\widetilde{\mathfrak{P}}_{\it surr}(B')=0$ and
$\widetilde{\mathfrak{P}}_{\it surr}(C')=1$, the proposition follows.
\hbox{}\hfill$\blacksquare$

\appendix \renewcommand{\theequation}{\thesection\arabic{equation}}
\setcounter{equation}{0}

\section{Hypergeometric Functions, P-Symbols}
\label{sec:hyper}

The following are facts about the generalized hypergeometric function
${}_{q+1}F_q$ and its ODE~\cite{Slater66}.  Let the rising factorial
$\alpha(\alpha+1)\cdots(\alpha+k-1)$ be denoted $(\alpha)_k$;
by~convention, $(\alpha)_0$~is interpreted as unity.  Then for any~$q\ge1$,
the function
${}_{q+1}F_q(\alpha_1,\ldots,\alpha_{q+1};\beta_1,\ldots,\beta_q;z)$ is
defined by
\begin{equation}
\label{eq:hyperseries}
{}_{q+1}F_q
\biggl(
\biggl.
\begin{array}{c}
{\alpha_1,\ldots,\alpha_{q+1}} \\
{\beta_1,\ldots,\beta_q}\\
\end{array}
\biggr| 
\,z
\biggr)
=
\sum_{k=0}^\infty
\frac
{(\alpha_1)_k\cdots(\alpha_{q+1})_k}
{(\beta_1)_k\cdots(\beta_{q})_k}
\,\frac{z^k}{k!}.
\end{equation}
Provided no denominator parameter~$\beta_i$ is a nonpositive integer, the
series coefficients are finite and the series converges absolutely on the
open unit disk $\left|z\right|<1$.  Provided $\Re
(\sum\beta_i-\sum\alpha_i) > 0$, it converges on~$|z|=1$, as~well.  Let
$\vartheta\defeq z\,d/dz$.  Then on the disk, ${}_{q+1}F_q$ satisfies the
order-$(q+1)$ ODE
\begin{equation}
\label{eq:hyperODE}
\Bigl[\vartheta(\vartheta+\beta_1-1)\cdots(\vartheta+\beta_q-1) -
z(\vartheta+\alpha_1)\cdots(\vartheta+\alpha_{q+1})\Bigr]F = 0.
\end{equation}
It~is the only solution analytic at~$z=0$
and equalling unity there.

The natural domain of definition of~(\ref{eq:hyperODE}) is the Riemann
sphere $\mathbb{CP}^1\defeq\mathbb{C}\cup\infty$.  By examination, this ODE
has $z=0,1,\infty$ as its only singular points, and is {\em Fuchsian\/}:
each singular point is regular.  ${}_{q+1}F_q$~can be continued to a
meromorphic function on $\mathbb{CP}^1\setminus\{0,1,\infty\}$, which is
generally multivalued.  In~fact, the solution space of~(\ref{eq:hyperODE})
is a $(q+1)$-dimensional space of multivalued meromorphic functions.
To~avoid multivaluedness, $\mathbb{CP}^1$~is cut along~$[1,\infty]$.
By~definition, ${}_{q+1}F_q$ is the continuation of the series from the
disk to $\mathbb{CP}^1\setminus[1,\infty]$.

The solution space of any order-$n$ Fuchsian ODE on~$\mathbb{CP}^1$ with
three singular points is determined to a large extent by their locations
and the $n$~characteristic exponents associated to each~\cite{Poole36}.
Let $z_1,z_2,z_3\in\mathbb{CP}^1$ denote the singular points, and
$\rho_1^{(i)},\ldots\rho_n^{(i)}\in\mathbb{C}$ the exponents of~$z=z_i$.
If $z_i\neq\infty$, this generally means that for each
$j\in\{1,\ldots,n\}$, the equation has a solution asymptotic to
$(z-z_i)^{\rho_j^{(i)}}$ as~$z\to z_i$.  (If~$z_i=\infty$, then
$(z-z_j)^{\rho_j^{(i)}}$ must be interpreted as $z^{-\rho_j^{(i)}}$.  Also,
if the difference between any pair of exponents of a singular point is an
integer, the solution corresponding to the smaller one may include a
logarithmic factor.)  This definition of characteristic exponents extends
immediately to ordinary points.  The $n$~exponents of any finite ordinary
point $z\neq z_1,z_2,z_3$ are $0,1,\ldots,n-1$.

The Riemann P-symbol for such an ODE, or for its solution space, is
\begin{equation}
P\left\{
\begin{array}{ccc|c}
z_1 & z_2 & z_3 & z \\
\hline
\rho^{(1)}_{1} & \rho^{(2)}_{1} & \rho^{(3)}_{1}  & \\
\vdots & \vdots & \vdots  & \\
\rho^{(1)}_{n} & \rho^{(2)}_{n} & \rho^{(3)}_{n}  & \\
\end{array}
\right\},
\end{equation}
where the order of exponents in each column is not significant.  This
tableau facilitates symbolic manipulation.  For example, multiplying the
general solution by $(z-z_0)^c$ will add~$c$ to the exponents of~$z=z_0$
and $-c$ to the exponents of~$z=\infty$.

It is readily verified that the ODE~(\ref{eq:hyperODE}) has exponents
$0,1-\beta_1,\ldots,{1-\beta_q}$ at~$z=0$, exponents
$0,1,2,\ldots,q-1,s$ at~$z=1$, and exponents
$\alpha_1,\ldots,\alpha_{q+1}$ at~$z=\infty$, where
$s\defeq\sum\beta_i-\sum\alpha_i$.  (This seems not to be well
known; it~is only partially explained in ref.~\citen{Slater66}.)  In~an
{\em ad~hoc\/} notation, we write
\begin{equation}
\label{eq:hyperPsymbol}
{}_{q+1}F_q
\biggl(
\begin{array}{c}
{\alpha_1,\ldots,\alpha_{q+1}} \\
{\beta_1,\ldots,\beta_q}\\
\end{array}
\!
\biggm|
z\,
\biggr)
\propto
P\left\{
{
\addtolength\extrarowheight{2.55pt}
\begin{array}{lll|l}
{0} & 1 & \infty & z \\
\hline
\fbox{0} & 0 & \alpha_1  & \\
1-\beta_1 & 1 & \alpha_2  & \\
\vdots & \vdots & \vdots  & \\
1-\beta_{q-1} & q-1 & \alpha_q  & \\
1-\beta_q & s & \alpha_{q+1}  & \\
\end{array}
}
\right\},
\end{equation}
with the box indicating that ${}_{q+1}F_q$ belongs to the zero exponent
at~$z=0$.  The sum of the $3(q+1)$ exponents equals ${{q+1}\choose2}$,
though this property is not specific to the hypergeometric ODE\null:
it~holds for any order-$(q+1)$ Fuchsian ODE on~$\mathbb{CP}^1$ with three
singular points.  So there are only $3q+2$ independent exponent parameters.

Any order-$(q+1)$ Fuchsian ODE on~$\mathbb{CP}^1$ with three singular
points, or more accurately its solution space, is characterized by the
$3q+2$ independent exponent parameters and ${q\choose2}$ additional
`accessory' parameters, which together with the exponent parameters
determine the global monodromy.  (See Poole~\cite{Poole36}, sec.~20, for a
normal form for the ODE from which the parameters may be extracted, with
some effort.)  The second-order (i.e.,~$q=1$) case is special in~that there
are no accessory parameters, and the solution space of the ODE is uniquely
determined by its P-symbol.  This is not the case when $q\ge2$, i.e., when
the ODE is of third or higher order.

If the $q+1$ exponents of one of the three singular points are
$0,1,\ldots,{q-1},s$ for some~$s$, up~to an overall additive constant, we
say the ODE is of {\em hypergeometric type\/}.  Provided its ${q\choose2}$
accessory parameters take suitable values, the solutions of any ODE of
hypergeometric type can be expressed in~terms of hypergeometric functions,
since it can be transformed to the hypergeometric ODE by redefining its
independent and dependent variables so as to~move its singular points to
$0,1,\infty$, and remove the additive constant.

\section*{Acknowledgements}
This work was partially supported by NSF grant PHY-0099484.

%\bibliographystyle{abbrv}
%\bibliography{general}

%\begin{bibliography}{99}
%\end{bibliography}

%\end{document}

\small\def\em{\it} \newcommand{\noopsort}[1]{} \newcommand{\printfirst}[2]{#1}
  \newcommand{\singleletter}[1]{#1} \newcommand{\switchargs}[2]{#2#1}

\end{document}